\documentclass[twocolumn,aps,prl,superscriptaddress]{revtex4-1}
\pdfoutput=1
\usepackage[colorlinks=true,citecolor=blue,linkcolor=blue,urlcolor=blue]{hyperref}
\usepackage{amsmath}
\usepackage{amssymb}
\usepackage{url,bbm,bm}
\usepackage{mathtools}
\usepackage{graphics,graphicx,color,float}

\usepackage{etoolbox} 
\usepackage{amsfonts}
\usepackage{amsthm}
\usepackage{tikz}
\usetikzlibrary{matrix}
\usepackage{xr-hyper}
\usepackage{tikz}
\usetikzlibrary{calc}

\newcommand{\beq}{\begin{equation}}
\newcommand{\enq}{\end{equation}}
\newcommand{\bel}{\begin{lemma}}
\newcommand{\enl}{\end{lemma}}
\newcommand{\bet}{\begin{theorem}}
\newcommand{\ent}{\end{theorem}}

\newcommand{\tr}{\mathrm{Tr}}
\makeatletter

\usepackage{booktabs,tabularx}

\newcommand{\suppress}[1]{}

\newcommand{\bra}[1]{\langle #1|}
\newcommand{\ket}[1]{|#1 \rangle}

\mathchardef\mhyphen="2D

\def\be{\begin{equation}}
\def\ee{\end{equation}}

\makeatletter
\newcommand*{\rom}[1]{\expandafter\@slowromancap\romannumeral #1@}
\makeatother

\makeatletter
\appto{\appendix}{%
 \@ifstar{\def\theequation@prefix{A.}}%
 {}%
}
\makeatother

\mathchardef\mhyphen="2D
\newtheorem{theorem}{Theorem}
\newtheorem{lemma}[theorem]{Lemma}

\newtheorem*{main result}{Main Theorem}

\newtheorem*{theorem*}{Theorem}

\begin{document}
	
\title{How to Teach AI to Play Bell Non-Local  Games: Reinforcement Learning }

\begin{abstract}
Motivated by the recent success of reinforcement learning in games such as Go and Dota2, we formulate Bell non-local games as a reinforcement learning problem. Such a formulation helps us to explore Bell non-locality in a   range of scenarios. The measurement settings and the quantum states are selected by the learner randomly in the beginning. Still, eventually, it starts understanding the underlying patterns and discovers an optimal (or near-optimal) quantum configuration corresponding to the task at hand. We provide a proof of principle approach to learning quantum configurations for violating various Bell inequalities. The algorithm also works for non-convex optimization problems where convex methods fail, thus offering a possibility to explore optimal quantum configurations for Bell inequalities corresponding to large quantum networks. We also implement a hybrid quantum-classical variational algorithm with reinforcement learning. \end{abstract}

\author{Kishor Bharti}
\affiliation{Centre for Quantum Technologies, National University of Singapore, 3 Science Drive 2, Singapore 117543, Singapore} 

\author{Tobias Haug}
\affiliation{Centre for Quantum Technologies, National University of Singapore, 3 Science Drive 2, Singapore 117543, Singapore}

\author{Vlatko Vedral}
\affiliation{Centre for Quantum Technologies, National University of Singapore, 3 Science Drive 2, Singapore 117543, Singapore} 
\affiliation{Clarendon Laboratory, University of Oxford, Parks Road, Oxford OX1 3PU, United Kingdom }

\author{Leong-Chuan Kwek}
\affiliation{Centre for Quantum Technologies, National University of Singapore, 3 Science Drive 2, Singapore 117543, Singapore} 
\affiliation{MajuLab, CNRS-UNS-NUS-NTU International Joint Research Unit, Singapore UMI 3654, Singapore}
\affiliation{National Institute of Education, Nanyang Technological University, Singapore 637616, Singapore}

\maketitle

\noindent {\em Introduction.---}In 2017, Alpha-Go defeated the best human Go player using tools from machine learning. Shortly after, the champion Alpha-Go program lost $100-0$ to its successor Alpha-Go Zero that was trained without human supervision \cite{silver2017mastering}. The machine learning tool responsible for the victory of Alpha-Go Zero is reinforcement learning (RL). RL  is one of the categories of machine learning, others being supervised and unsupervised learning \cite{sutton2018reinforcement, shalev2014understanding}. Aside from Go, RL has been successfully applied in many other games including Atari and multiplayer games like Dota2 \cite{mnih2015human,Dota}.

Recently, the synergy between machine learning and quantum information has led to some impressive fundamental progress at their interface, ranging from many-body physics to quantum feedback and control \cite{Carleo602,bukov2018reinforcement, niu2019universal, fosel2018reinforcement, melnikov2018active, dunjko2016quantum, hentschel2010machine}. 
Moved by the success of reinforcement learning, it is natural to inquire if one can use the tools from reinforcement learning to play quantum games. In this Letter, we employ reinforcement learning techniques for playing Bell non-local games.

\noindent
Experiments conducted on spacelike separated entangled systems lead to correlations not allowed by any local realistic theory and lead to violation of Bell inequalities \cite{Bell64}.
Apart from its fundamental significance in understanding the nonclassical nature of the physical world, Bell nonlocality has found practical applications in emerging quantum technologies like quantum key distribution (QKD), randomness certification and self-testing \cite{ekert1991quantum, pironio2010random, Yao_self}.

\par
\noindent
The essence of Bell inequality violation can be captured via the winning strategies for the multi-player co-operative games called called Bell non-local games \cite{cleve2004consequences,condon1989complexity}. The connection between Bell inequalities and Bell non-local games is deciphered from the simple observation that questions are nothing but labels for measurement settings. One prominent example of Bell non-local game is Clauser-Horne-Shimony-Holt (CHSH) game \cite{CHSH,cleve2004consequences}.  
\par 
\noindent
The winning probabilities for Bell non-local games depend on the joint strategy chosen by the players. Such strategies are often captured via some local hidden variable model (classical model)  or some quantum model (sharing quantum resources). Discovering a quantum strategy which outperforms any classical strategy leads to the confirmation for Bell non-local character of quantum theory. A quantum strategy corresponds to finding a set of local measurement settings for every party and some entangled state. The task, as mentioned above, is a sequential decision-making problem and hence, can be tackled via reinforcement learning. Since reinforcement learning has already been applied to complex decision-making problems such as playing the game of Go, it motivates us to use RL for Bell non-local games.
\par
\noindent
Techniques from reinforcement learning have turned out to be useful in a wide range of problems in quantum physics which involve sequential decision making. The goal of reinforcement learning is to convert the optimization problem to a set of policies which dictate the behaviour of an AI agent (we will refer to such an agent as RL agent). The RL agent learns to choose actions from the set of actions available at each stage of the sequential decision-making process based on the reward it receives. The goal of the AI is to find a sequence of actions such that the reward is maximized. Such a strategy betters the RL agent to find the optimum (or almost optimum) solution to the problem at hand.
\par
\noindent
There have been several works that combine the advantages of machine learning and quantum foundations \cite{deng2018machine,canabarro2019machine, krivachy2019neural}. In \cite{canabarro2019machine}, authors employed machine learning to distinguish between classical, quantum, and post-quantum correlations using supervised learning. A solution to the Membership problem (local versus non-local) for causal structures was proposed in \cite{krivachy2019neural} via neural networks. In \cite{deng2018machine}, finding the maximum value of a many-body Bell inequality for the case of fixed measurement settings was tackled using reinforcement learning. The approach in \cite{deng2018machine} depends on the mapping between finding the ground state of a Hamiltonian to obtain the largest eigenvalue of a Bell operator. However, this scheme  works only for convex scenarios.
\par
\noindent
The measurement settings and the quantum state can be specified using finitely many parameters which can be tuned by the RL agent via training to find the optimum (or near optimum) configuration. The performance of machine learning algorithms depends on the representational ability of the machine learning model, as well as the optimization protocol. A meticulously chosen representation can help the RL agent to learn better. Motivated by the success of variational circuit on near term hardware, we demonstrate a use-case of such circuits to parametrize our quantum state. The purpose of the RL agent remains to tune the parameters to get optimum performance.
\par
\noindent {\em Background and main idea.---} According to Bell's theorem, experiments conducted on spacelike distributed entangled quantum systems refute the local realism assumption. The most simplistic experimental scenario which can testify the violation of local realism requires two spacelike separated parties with two dichotomic measurement settings on each side. The scenario involving $n$ spacelike separated parties with $m$ $k$-outcome measurements on each side is denoted by $(n,m,k)$ scenario.  The scenario, as presented above, is $(2,2,2)$ scenario. The probability distribution $p(a,b\vert x,y)$ for measurement settings $x,y$ and respective outcomes $a,b$ if describable by a hidden variable model admits the following description,
$$
p(a,b \vert x,y) = \sum_{\lambda} p(a\vert x,\lambda) p(b \vert y, \lambda) p(\lambda).
$$
The probabilities of the above kind in quantum theory admit following representation,
\be
p(a,b \vert x,y) = \tr \left[ \left(M_{x}^{a}\otimes M_{y}^{b}\right)\rho_{AB}\right],
\ee
for some joint density matrix $\rho_{AB}$ and measurement operators $M_x^a$ and $M_y^b$. The linear and polynomial Bell inequalities consist of linear/polynomial combinations of probabilities of the above form. The set of all the probabilities ${\bf{P}} = \{p(a,b \vert x,y) \}$ is called Behaviour. If a particular selection of quantum measurement settings and state violate the bound on the Bell inequality, they are said to violate local realism or in other words demonstrate Bell non-local nature of quantum theory. The task of discovering the measurement settings and state which lead to the maximum violation can be a convex \cite{CHSH, brunner2014bell} or a non-convex optimization program \cite{chaves2016polynomial, tavakoli2014nonlocal}. 
\par
\begin{figure}[htbp]
	\centering
	\includegraphics[width=0.35\textwidth]{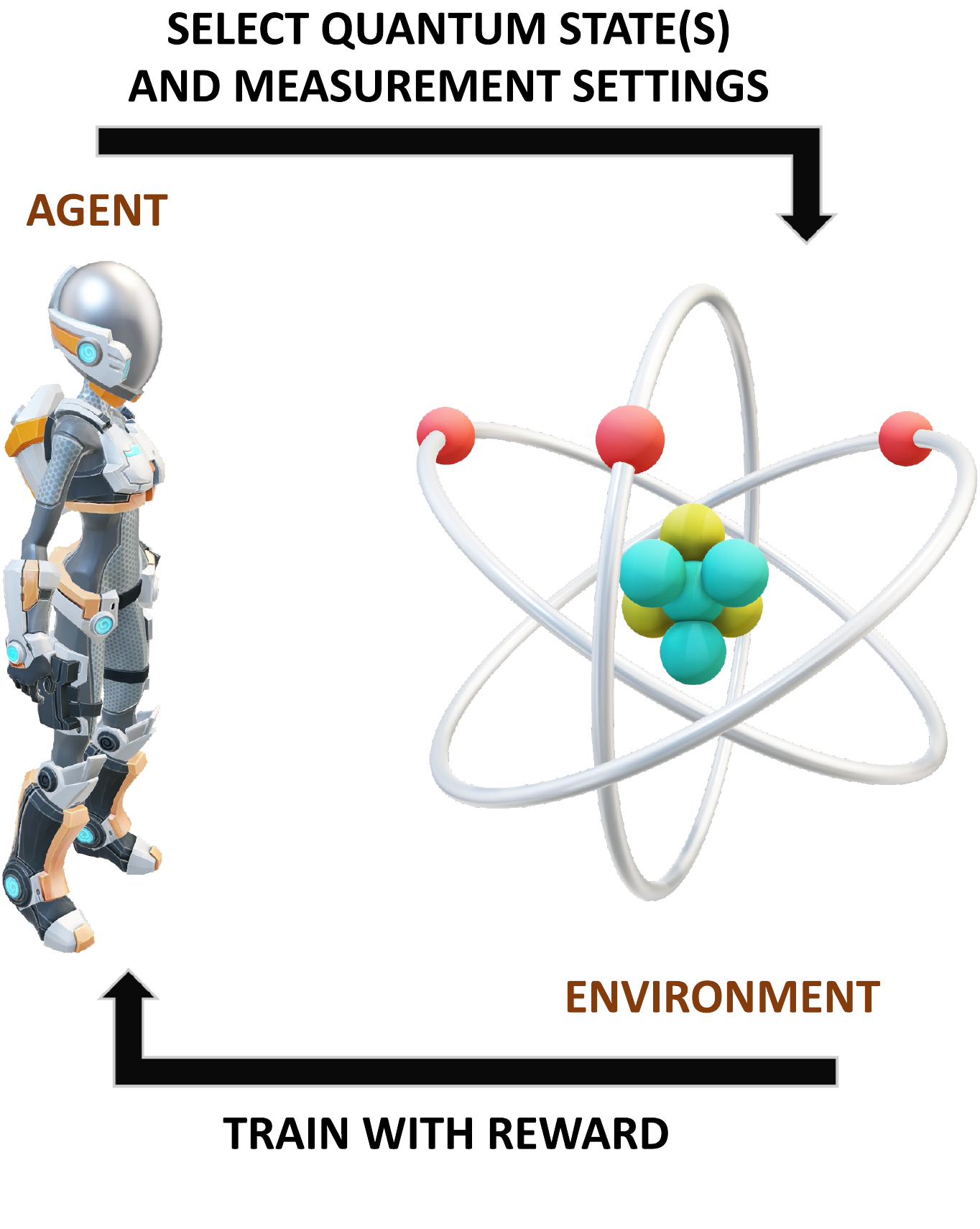} 
	\caption{The above cartoon depicts our approach to find the best measurement settings and quantum state corresponding to maximum quantum violation of a Bell inequality in a fixed dimension. For example in the case of CHSH inequality, the agent selects Alice's and Bob's measurement settings plus the entangled quantum state i.e $A_1, A_2, B_1, B_2$ and $\rho = \ket{\psi}\bra{\psi}$.  The reward is given at the end with $R = \tr ({\cal{B}} \rho)$ where ${\cal{B}} = A1\otimes B1 +A1\otimes B2 +A2\otimes B1 -A2\otimes B2$ and used to train the agent.}
	\label{fig:Policy_Ag_env_CHSH}
\end{figure}
\noindent
The task of selecting an optimum configuration can be framed as a sequential decision-making process. The Bell non-local games corresponding to the Bell inequalities involve finding measurement settings and sharing quantum state such that they win the game. We train our RL agent to find measurement settings and state (see Fig.\ref{fig:Policy_Ag_env_CHSH}). The CHSH inequality is the only non-trivial facet-defining Bell inequality for $(2,2,2)$ scenario (also known as Bell-CHSH scenario)  and thus captures Bell non-locality for the smallest possible scenario.  Let $\langle a_x b_y \rangle = \sum_{a,b}  ab p(a,b\vert x,y)$ for $x,y \in \{1,2\}$ and $a, b \in \{-1, +1\}$. The CHSH inequality is given by 
\be
\langle a_1 b_1 \rangle + \langle a_1 b_2 \rangle  + \langle a_2 b_1 \rangle  - \langle a_2 b_2 \rangle  \leq 2.
\ee
For suitably chosen quantum measurements and state, the above inequality can be violated up to $2 \sqrt{2}.$ We discuss multiple approaches to recast Bell scenario as a game with the Bell-CHSH scenario as the test-bed.
\par
\noindent%
(a) Fixed state approach:
In this approach, one fixes the quantum state and vary the measurement settings. The procedure can be summarized as follows:
\begin {enumerate}
\item Fix the quantum state, call it $\rho$.
\item \begin{enumerate}
\item Alice selects two measurement operators from her Hilbert space, denoted by  $A1 \text{ and } {A2} \in \cal{H_A}$. Since the Hilbert spaces for Alice and Bob are different, the action of these measurement operators on Bob's Hilbert space is equivalent to identity.
\item Bob selects two measurement operators from his Hilbert space, denoted by  $B1 \text{ and } {B2} \in \cal{H_B}$.
\end{enumerate}
\item The reward function is given by $R = \tr ({\cal{B}} \rho)$ where ${\cal{B}} = A1\otimes B1 +A1\otimes B2 +A2\otimes B1 -A2\otimes B2$.
\end {enumerate}
The goal is to maximize the reward (given in step $3$) for different choices corresponding to step $2$. 
This approach can be explored by fixing the quantum state as a maximally entangled state or some experimentally accessible state. However, since there is no freedom in choosing the state, the overall strategy will be often sub-optimal.

\par
\noindent %
(b) Exact diagonalization approach:
In this approach, we do not fix the quantum state. In fact, we find the optimal quantum state for a given measurement setting via exact diagonalization techniques:
\begin {enumerate}
\item Same as the step $2$ in the fixed state approach.
\item The reward function is given by $R = eigmin(- {\cal{B}})$ where ${\cal{B}} = A1\otimes B1 +A1\otimes B2 +A2\otimes B1 -A2\otimes B2$ and $eigmin$ refers to the minimum eigenvalue. \end {enumerate}
The goal is to maximize the reward (given in step $2$) for different choices corresponding to step $1$. The upside of this approach is that we get optimal measurement settings and the state (both) for the given scenario. However, the downside is that exact diagonalization can be challenging for complex scenarios.

\par
\noindent %
(c) Full RL approach:
In this approach, we vary both the quantum state and the measurement settings:
\begin {enumerate}
\item Same as the step $2$ in the fixed state approach.
\item The parameters representing the density matrix ($\rho$) is selected from an appropriate range.
\item The reward function is given by $R = \tr ({\cal{B}} \rho)$ where ${\cal{B}} = A1\otimes B1 +A1\otimes B2 +A2\otimes B1 -A2\otimes B2$.
\end {enumerate}

To train the RL agent, we employ the proximal policy optimization (PPO) algorithm, which is a state of the art RL algorithm \cite{schulman2017proximal}, using the OpenAI Spinning Up implementation \cite{spinningup} (see Section \ref{app_PPO} in the Appendix for further details).

\noindent {\em Test Cases.---}
Here, we implement reinforcement learning on a variety of test cases.
\vspace{0.1cm}
\par
\begin{figure}[h]
\centering
\includegraphics[width=0.5\textwidth]{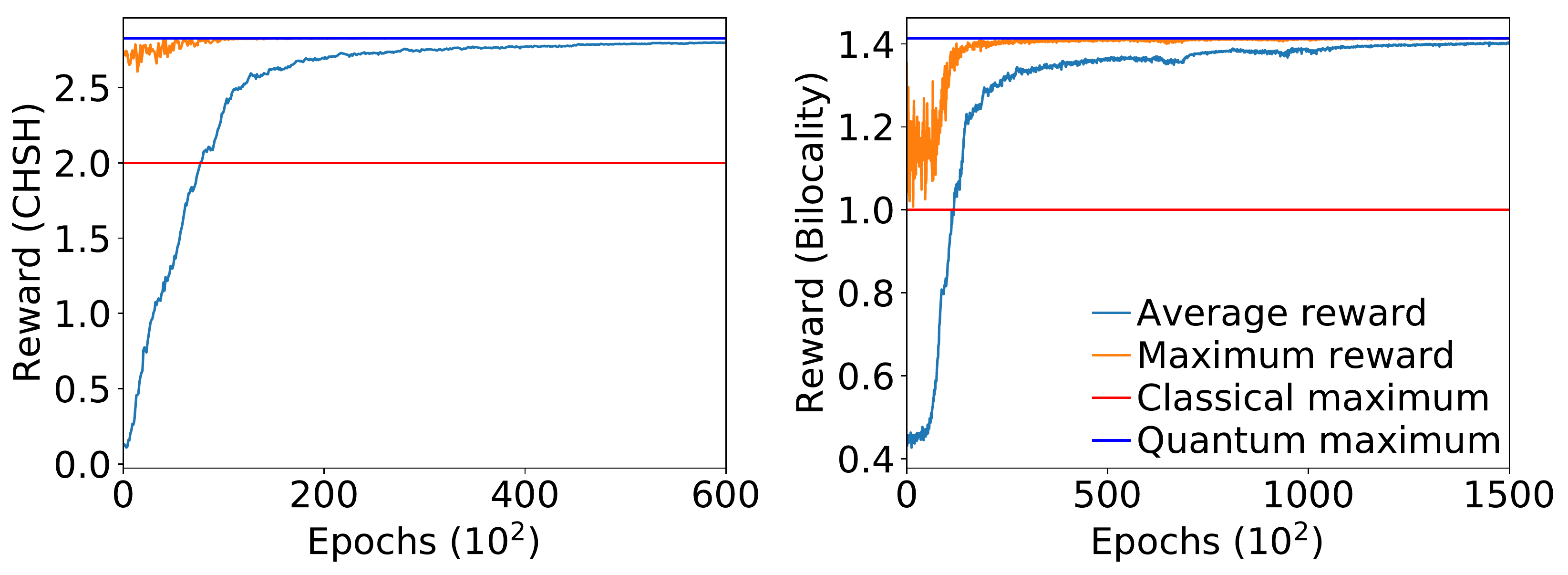} 
\caption{CHSH and bilocality inequality. Left: We plot the results for CHSH inequality. The classical and the quantum bounds on the inequality are $2$ and $2\sqrt{2}$ respectively. The agent randomly generates measurement settings and state in the beginning. After a few training epochs, the maximum performance within a epoch goes to the quantum maximum. The average performance also improves over epochs. In the beginning, the average performance is below the classical maximum, but eventually approaches the quantum maximum.\\
Right: Unlike the CHSH case, the task of finding the quantum maximum is a non-convex optimization program. The classical and the quantum maximum for the bilocality inequality are $1$ and $\sqrt{2}$ respectively. Over a few epochs, the maximum performance approaches the quantum maximum. Similar behaviour is observed for the average performance.}
\label{fig:CHSH_Bil}
\end{figure}

\noindent
{{(a) CHSH inequality:}} We train our agent to find qubit positive-operator valued measures (POVM) corresponding to Alice's/Bob's measurement operators and two-qubit quantum state which correspond to the optimal quantum violation. Refer to the left panel of Fig. \ref{fig:CHSH_Bil} for details.

\vspace{0.1cm}
\par
\noindent
{{(b) Bilocality inequality: }}The Bilocality inequality captures the Bell nonlocality of three spacelike separated parties, say Alice, Bob and Charlie interconnected by two independent sources of states. The independence of sources is known as bilocality assumption i.e., $p\left(\lambda_1, \lambda_2 \right) = p\left(\lambda_1\right)p\left(\lambda_2\right) $, which forces the tripartite observed statistics to have the following form,
\begin{align*}
p(a,b,c \vert x,y,z) = \sum_{\lambda_1,\lambda_2} p\left(a \vert x, \lambda_1 \right) p\left(b \vert y, \lambda_1, \lambda_2 \right)p\left(c \vert z, \lambda_2 \right) \\
 p\left(\lambda_1\right)p\left(\lambda_2\right).
\end{align*}
For tripartite bilocality scenario with two measurement choices per party $(x,y,z = 0,1)$, one of the inequalities is given by  
\be
\sqrt{\vert I \vert } + \sqrt{\vert J \vert} \leq 1,
\ee
where $I = \left( 1 \over 4 \right)\sum_{x,z}\langle A_x B_0 C_z \rangle$ and $J = \left( 1 \over 4 \right)\sum_{x,z}(-1)^{x+z}\langle A_x B_1 C_z \rangle$. The best quantum value known till date is $\sqrt{2}$ \cite{tavakoli2014nonlocal}. The set of quantum behaviours for the bilocality scenario is non-convex. In \cite{tavakoli2014nonlocal}, authors fix the measurement settings for Alice and Charlie as well the the overall quantum state, and optimize to find Bob's measurement settings. We do the opposite i.e fix Bob's settings and ask the RL agent to learn everything else. Refer to the right panel of Fig. \ref{fig:CHSH_Bil} for details.
\vspace{0.1cm}
\par
\noindent
{(c) Many body symmetric Bell inequality: } Now we consider an N-partite Bell inequality where each party can select one of the two dichotomic (outcome $\pm 1$ here) measurements. Following the convention in \cite{tura2014detecting}, we denote the measurements by ${\cal{M}}_{j}^{(i)}$ for $j =0,1$ and $i = 1,2,\cdots, N$. The one and two body symmetrized correlators with respect to  ${\cal{M}}_{j}^{(i)}$ are given by
\be
\label{Correlators}
S_k = \sum_{i=1}^{N} \langle {\cal{M}}_{k}^{(i)}\rangle \quad \text{and} \quad S_{kl} = \sum_{{i,j=1}_{(i \neq j)}}^{N} \langle {\cal{M}}_{k}^{(i)}{\cal{M}}_{l}^{(j)}\rangle.
\ee
\par
\begin{figure}[h]
\centering
\includegraphics[width=0.5\textwidth]{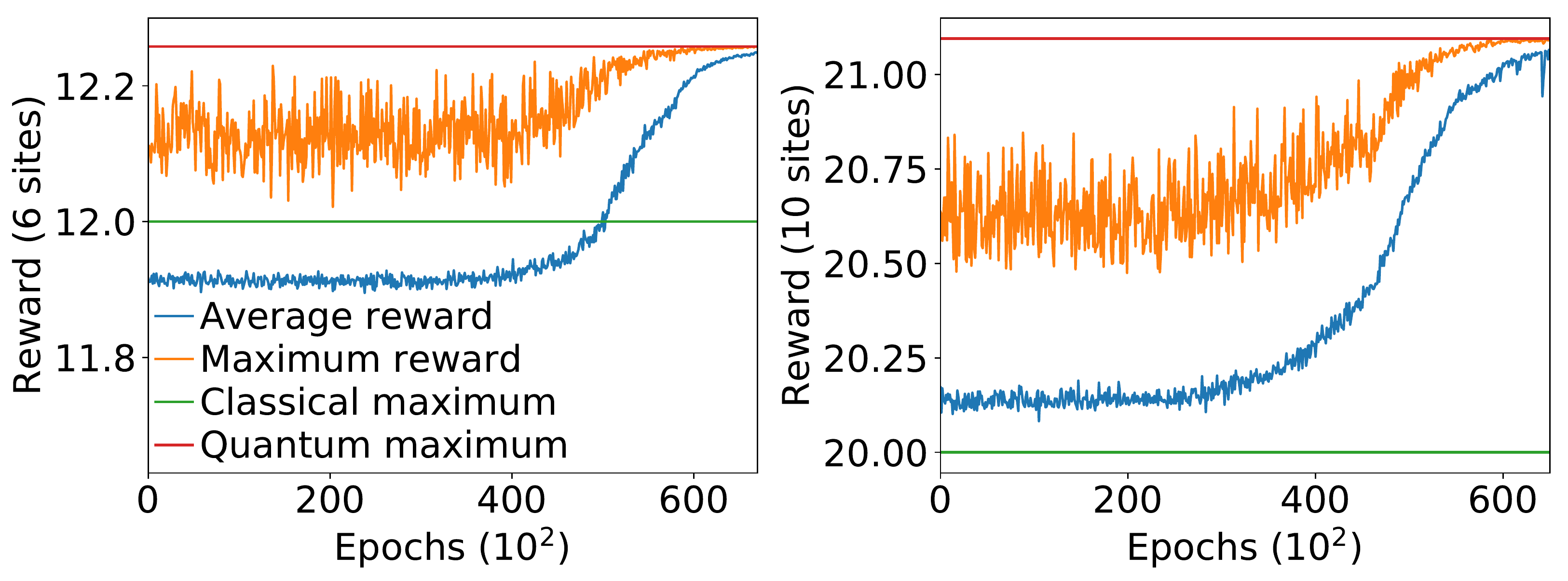} 
\caption{Many-body Bell inequality. Left: We train the RL agent to find measurement angles for maximal quantum violation for the inequality in \eqref{MBI} for $N=6$ qubits using exact diagonalization In the beginning, the average performance is less than the classical bound. Both the average as well as maximum performance improves with training and reaches to the optimum value. \\
Right: We present the plot for $N=10$ case. Unlike the previous case, exact diagonalization approach helps the agent perform non-classically from the beginning. Overall we see similar behaviour as $N=6$ case.}
\label{fig:MB_10}
\end{figure}

For the above scenario, we employ the following Bell inequality
\be
\label{MBI}
-2S_0 + {1 \over 2} S_{00} - S_{01} + {1 \over 2}S_{11} + 2N \geq 0.
\ee
for $N =6$ and $10$ sites (see Fig. \ref{fig:MB_10}) . The RL agent achieves the optimal quantum violation (as computed in \cite{tura2014detecting}) after training for a while.

\vspace{0.1cm}
\par
\noindent
\par
\noindent
{(d) Variational quantum-classical hybrid case: Finally, we train our agent corresponding to another type of many-body Bell inequalities. Let us take the following Bell inequality
\be
\alpha S_0 + \beta S_1 + {\gamma \over 2} S_{00} + \delta S_{01} + {\epsilon \over 2} S_{11} \geq -\beta_{c}.
\label{Dicke}
\ee
Let us denote ceiling of $(N-1) \over 2$ by $C_N$ and the floor by $F_N$. Following \cite{tura2014detecting}, we fix  $\alpha_N = N(N-1)(C_N - {N \over 2})$, $\beta_N = {\alpha_N \over N}$, $\gamma_N = {N(N-1) \over 2}$, $\delta_N = {N \over 2}$ and $\epsilon_N = -1$. For this choice of parameters, the classical bound on the Bell inequality is given by $\beta_C^{N} = {1 \over 2}N(N-1)C_N$. This inequality is violated for Dicke states, which are of experimental relevance~\cite{tura2014detecting}. We train the RL agent to select  measurement settings and quantum states such that it violates the classical bound. We provide it a variational circuit with tunable parameters. We start from a product state of all qubits in state zero. A single layer of our $N$ qubit variational Ansatz consists of  $N$ tunable local rotations around $Y$ axis, followed by CNOT gates in a ring fashion. After a $k$-layer circuit, we implement last layer of rotations. This Ansatz is also known as hardware-efficient Ansatz \cite{kandala2017hardware}. For training details, refer to Fig. \ref{fig:Var_dicke}. Indeed, we find that for sufficient amount of layers the violation can reach the quantum maximal bound.
\begin{figure}[h]
\centering
\includegraphics[width=0.5\textwidth]{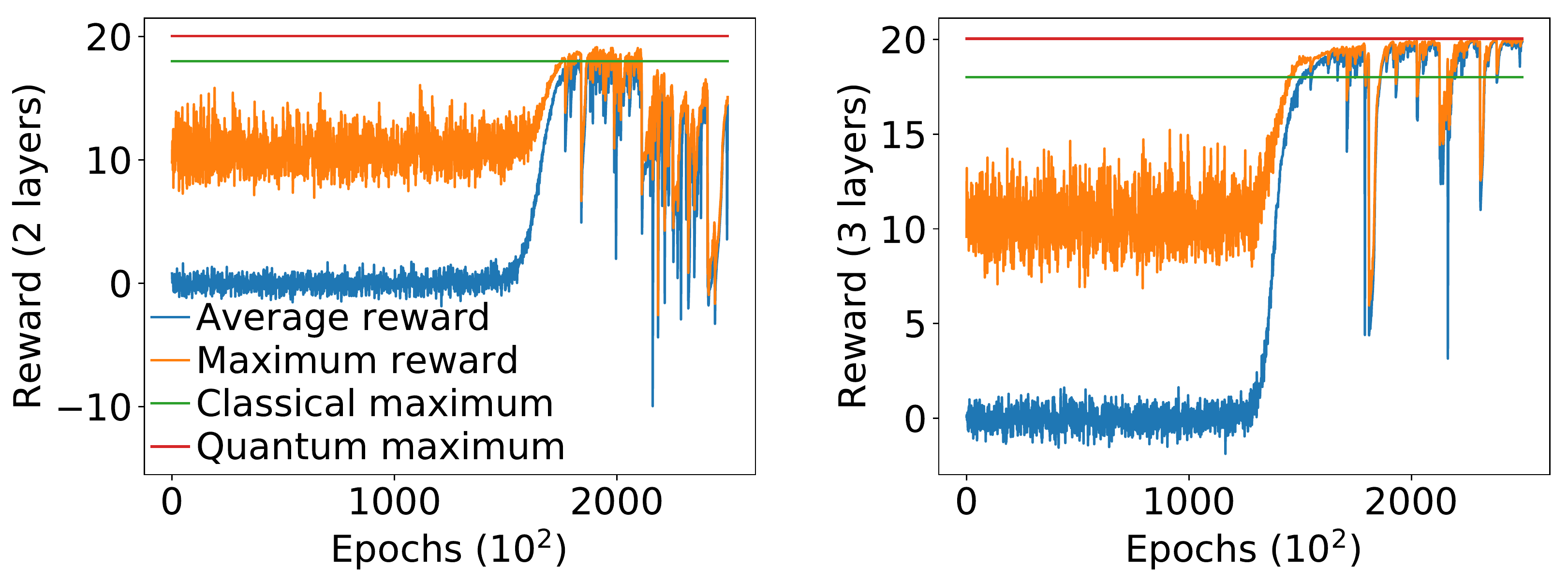} 
\caption{Variational classic-quantum hybrid case. Left: We train the RL agent to violate the Bell inequality in \eqref{Dicke}. The quantum state is selected by tuning the parameters of an $N$-qubit variational circuit. A two layered circuit barely violates the classical bound. \\ Right: We observe the violation up to quantum maximum by increasing the number of layers to three. With increasing layer depth, the range of states that can be reached increases, thus allowing us to reach a state that is close to maximal violation of the inequality.}
\label{fig:Var_dicke}
\end{figure}
\par 
\noindent {\em Discussions and Conclusions.---}
Finding the set of angles or quantum states that violate Bell inequalities can be a challenging problem, especially in a non-convex setting or for hybrid quantum-classical approaches.  Reinforcement learning provides a new approach to understand and optimize Bell games. State-of-the-art method like proximal policy optimization allow to treat this problem from the perspective of a sequential decision process. 
For our current approach, the agent is given a reward at the end of every epoch, which is equal to the expectation value of the Bell operator for the selected configuration of state and measurements. Such a reward scheme is sparse and might not be scalable. A potential future direction should be to develop better reward functions.
\begin{figure}[h]
\centering
\includegraphics[width=0.5\textwidth]{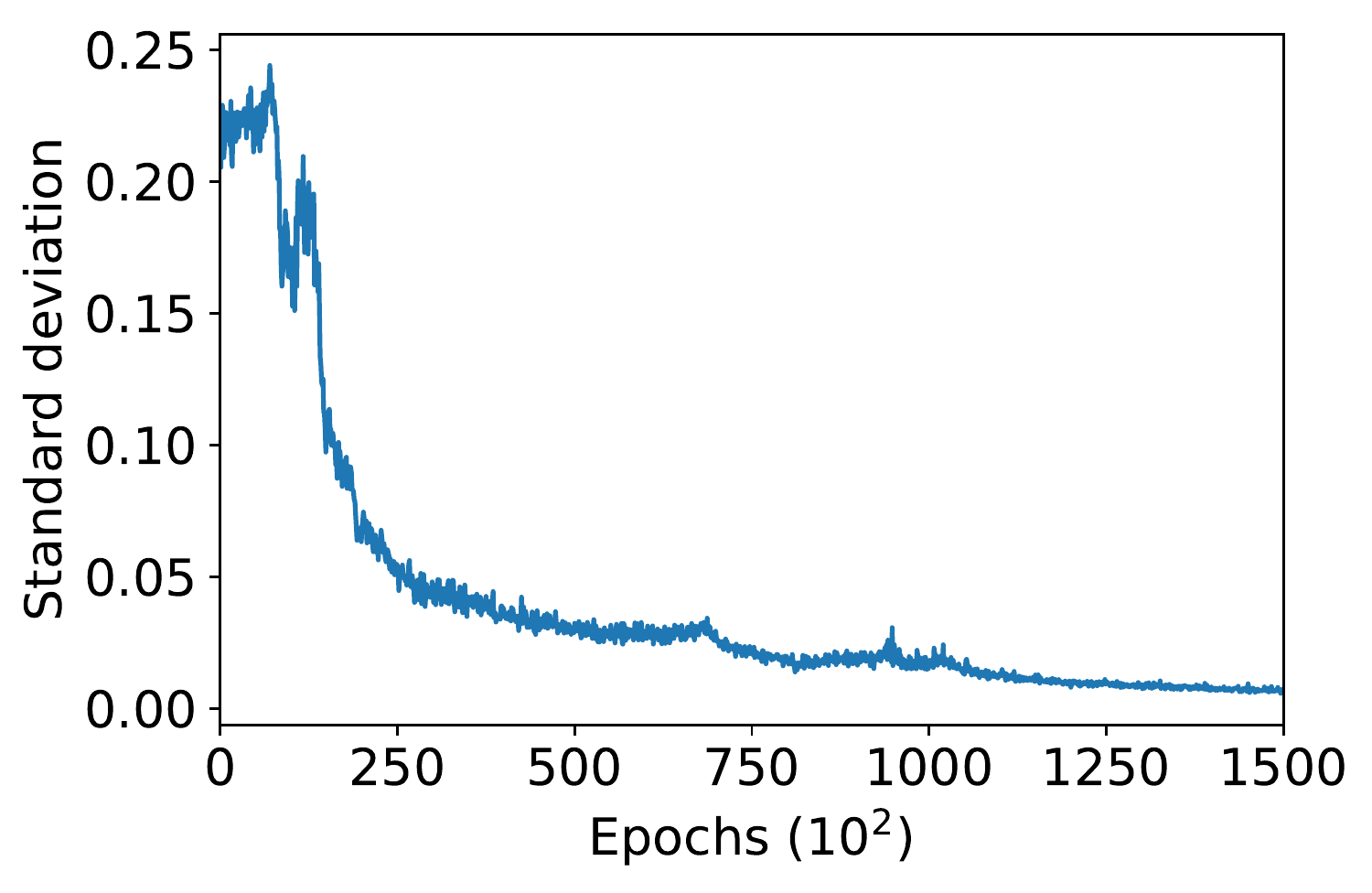} 
\caption{The above plot shows the drop in standard deviation of the reward as the training for RL agent proceeds. With prolonged training the agent tends to exploit the best available option and explores less. Eventually, the standard deviation approaches zero.}
\label{fig:Std}
\end{figure}
\noindent
\par
One of the main advantages of our reinforcement learning approach concerns the handling of the trade-off between exploration vs exploitation. At the beginning of the training, the RL agent tries to explore the many possibilities. Such an exploration leads to a large standard deviation in the reward, which goes to zero as the agent learns to exploit the nearly optimal choice of possibilities (see Fig. \ref{fig:Std}). This procedure could allow to optimize even very complex landscapes and find optimal or sub-optimal solutions in settings where other methods have failed so far.
\par
\noindent
The success of RL for real-world decision-making problems provides confidence for their success in tackling non-convex Bell scenarios, most straightforward of which is the bilocality scenario. Bell scenarios corresponding to large quantum networks is highly non-convex, which renders conventional computational tools based on semidefinite programming and other convex optimization algorithms not very helpful. We believe that state of the art reinforcement learning methods such as PPO could be useful to undertake problems in Bell scenarios corresponding to networks.
\noindent
\par
In this work, we trained a single agent learning the optimization landscape for various Bell scenarios to find optimum measurement settings and state for all the parties. A possible future direction would be to treat every party as an agent and learn the landscape using multi-agent reinforcement learning \cite{lowe2017multi}.
\par
\noindent
We represented our quantum state as a variational circuit with free parameters which can be tuned by the RL agent. Such variational circuits are a promising approach to harness the potential of Noisy Intermediate-Scale Quantum (NISQ) devices. Thus, integration of RL with variational hybrid classical-quantum optimization is a possible future direction worth investigating \cite{preskill2018quantum}.
\par 
\noindent {\em Acknowledgements.---}
The authors are grateful to the National Research Foundation and the Ministry of Education, Singapore, for financial support. We thank Valerio Scarani for discussions.

\bibliography{RL_Bell}
\appendix
\section{Bell games and Bell inequalities}
The essence of Bell inequalities is often captured via Bell non-local games. Such games have been studied in detail by computer scientists in the context of entanglement assisted interactive proof systems \cite{cleve2004consequences}. A typical example of Bell non-local game (or Bell game in short) consists of a referee who plays the games against parties say, Alice and Bob. The parties or players are known as provers. The referee is known as the verifier. The verifier selects questions $x \in X$ and $y \in Y$ for Alice and Bob from some probability distribution $\pi: X \times Y \rightarrow [0,1]$. The provers return answers  $a \in R_A$ and $b \in R_B$. The verifier uses a predicate $V: R_A \times R_B  \times X \times Y \rightarrow \{0,1\}$ such that $V(a,b,x,y) = 1$ whenever the provers win against the verifier by providing answers $a$ and $b$ for questions $x$ and $y$. Any strategy selected by the provers lead to probabilities $p(a,b \vert x,y)$ for questions $x,y$ and answers $a,b$. The winning probability for the game is given by
\be
p_{win} = \sum_{x,y} \pi(x,y) \sum_{a,b} V(a,b \vert x,y) p(a,b \vert x,y).
\ee
The connection between Bell inequalities and Bell non-local games can be deciphered from the simple observation that questions are nothing but labels for measurement settings. The maximization of winning probabilities over all deterministic strategies correspond to Bell inequalities.
\section{Markov decision process}
The mathematics behind reinforcement learning can be captured via the Markov decision process. A Markov decision process (MDP) is a powerful tool to capture the mathematics of decision making where a subset of outcomes are random. Thus the decision-maker has partial control on the remaining outcomes only. An MDP is characterized by a five-tuple, $\{S,A,R,P, \rho_0\}$, where the meaning of the respective symbols is as follows.
\begin{itemize}
\item $S$: It denotes the set of all possible states. The state at time-step `t' is denoted by $s_t$.
\item $A$: It represents the set of all possible actions. The state at time-step `t' is denoted by $a_t$.
\item $R$ : The reward function $R$ is a map $S \times A \times S \rightarrow \mathbb{R}$ where $\mathbb{R}$ denotes the set of real numbers. In particular, the reward the time-step `t' is given by $r_t = R\left(s_t, a_t \right)$.
\item $P$ : The transition probability $P$ is a map $S \times A \rightarrow \mathcal{P}(S) $. Quite specifically, the transition probability $P\left(s'\vert s, a\right)$ denotes the probability of transitioning to the state $s'$ if the current state is $s$ and action taken is $a$.
\item $\rho_0$: It denotes the initial state distribution.
\end{itemize}
The ``Markov" in MDP refers to the Markov property which asserts that the transitions depend only on the current state of the system and action.
\section{Reinforcement learning}
Reinforcement learning is one of the branches of machine learning, which involves the study of agents who learn via trial and error. The learning happens via rewarding (punishing)  the agent for desirable (unwanted) actions. The successful training in reinforcement learning involves a balance between exploration of unknown landscape with the exploitation of the prior experience. The training of an RL agent involves agent-environment interaction. The agent perceives the state of the environment and takes action. Due to the agent's action, the state of the environment changes and the agent receives some reward or penalty. The rule-book used by the agent to select an action from the set of actions (known as action space) is called policy. The goal of reinforcement learning is to discover the optimal policy. 
For a pictorial understanding of the setting, the reader can refer to Fig. \ref{fig:Policy_Ag_env}.
\begin{figure}[H]
\centering
\includegraphics[width=0.45\textwidth]{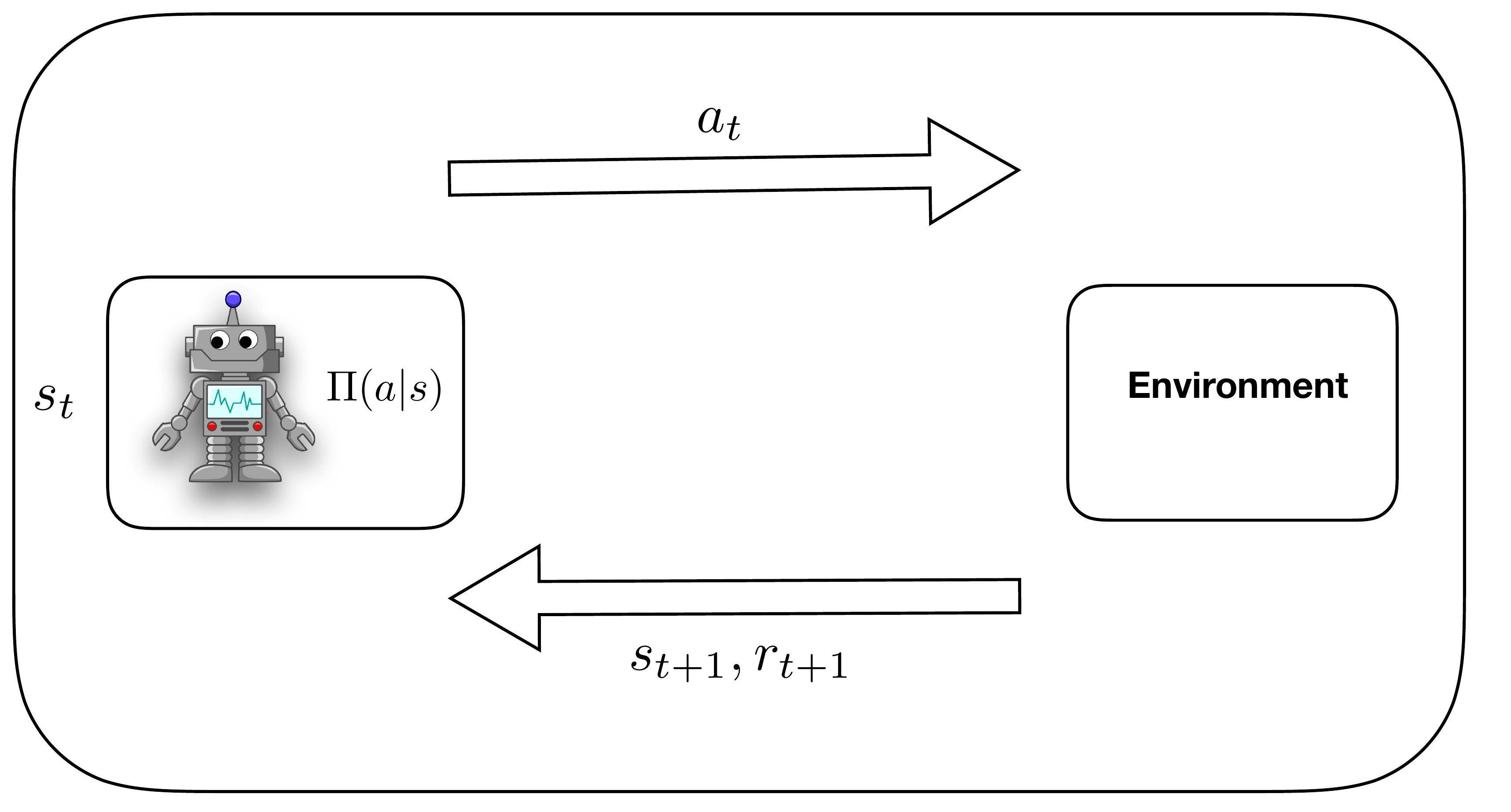} 
\caption{The above cartoon describes the interaction of an agent with the environment. The agent perceives the state of the environment as $s_t$ and takes action $a_t$ by following a policy $\Pi$. The state of the environment shifts to $s_{t+1}$ and the agent receives the reward $r_{t+1}$. The goal of the RL agent is to find the optimal policy, which leads to the highest expected total reward for a finite number of agent-environment interactions.}
\label{fig:Policy_Ag_env}
\end{figure}
\par
The state of the environment at the onset of agent-environment interaction round $t$ will be denoted by $s_t$. The set of all valid actions which an agent can take is called action space. The action space can be discrete or continuous. The action taken by the agent in round $t$ will be denoted by $a_t$. Once the agent performs the action $a_t$, the state of the environment changes to $s_{t+1}$ and the agent gets reward $r_t$. Agent-environment interaction leads to a series of states and actions. Such a series is usually called trajectory, episode or rollout. We will denote the trajectories by $\tau$. A trajectory $\tau$ looks like $\{s_1,u_1,s_2,u_2,\cdots,s_H,u_H\}$. The maximum number of interactions (say $H$) sets the length of the trajectory and is known as Horizon. For our purposes, we will work with finite horizons. As an agent follows a trajectory, it reaps the rewards via its interaction with the environment. Therefore, one can associate a total reward value to a trajectory. Let $P(\tau)$ be the probability of a particular trajectory $\tau$ and the corresponding total reward is $R(\tau)$.
The expected reward is given by $\sum_{\tau} P(\tau) R(\tau).$ In RL, it is often useful to estimate the ``value'' of a state or a state-action pair. It gives us an estimate of the expected reward if one starts in the state or state-action pair and follows a particular policy. Due to the aforementioned need, a few value functions have been defined in the literature.
\begin{enumerate}
\item On-policy value function $\left( V^{\pi}(s) \right)$: It quantifies the expected return if the RL agent starts in the state $s$ and follows the policy $\pi$ forever.
\item On-policy action-value function $\left( Q^{\pi}(s,a) \right)$: It gives the expected return if the RL agent starts in the state $s$, takes an arbitrary action $a$ and then follows the policy $\pi$ forever.
\item Optimal value function $\left( V^{\star}(s) \right)$: It quantifies the expected return if the RL agent begins in the state $s$ and follows the optimal policy $\pi^{\star}$ permanently.
\item Optimal action-value function $\left( Q^{\star}(s,a) \right)$: It gives the expected return if the RL agent begins in the state $s$, takes an arbitrary action $a$ and then follows the optimal policy $\pi^{\star}$ forever.
\end{enumerate}
\section{Policy gradient}
In this section, we will review the mathematics of policy gradient. The policy tells the agent to determine which action to take depending on the state of the environment. 
A policy $\Pi$ takes state $s_t$ as input and outputs probability distribution over all possible actions. We denote the parameters which characterize a policy $\Pi_{\theta}$ by $\theta$. For an agent following the policy $\Pi_{\theta}$, the expected reward is given by 
\be
J(\theta) = \sum_{\tau} \Pi_\theta(\tau) R(\tau).
\ee
The goal of policy gradient is to find the optimum policy $\Pi_{\theta^{\star}}$ such that $J\left(\theta^{\star}\right)$ is maximum over $\theta$. In other words, the target is to model a policy that creates trajectories which maximize the total rewards. A possible approach to do so is by taking the gradient of $\Pi_{\theta}(\tau)$ for $\theta$ and do gradient ascent. Since the gradient is over the space of policies, the algorithm to find the optimum policy via this approach is called policy gradient, justifying its name. Since
\begin{align*}
J(\theta) = \int \Pi_{\theta}(\tau) R(\tau) d\tau,
\end{align*}
taking its gradient with respect to $\theta$, we get 
\begin{align*}
\nabla_{\theta}J(\theta) = \int \nabla_{\theta} \Pi_{\theta}(\tau) R(\tau) d\tau \\
= \int \Pi_{\theta} \nabla_{\theta} \log \left(\Pi_{\theta}(\tau)\right) R(\tau) d\tau \\
= \mathbb{E}_{\tau \sim \Pi_{\theta}(\tau)} \left[ \nabla_{\theta} \log \left(\Pi_{\theta}(\tau)\right) R(\tau) \right],
\end{align*}
where $\mathbb{E}_{\tau \sim \Pi_{\theta}}$ denotes expectation over trajectories $\tau$ with probability determined by the policy $\Pi_{\theta}$.
Since the policy gradient can be represented as expectation, we can use sampling to approximate it. Thus, the policy gradient algorithm reduces to
\begin{align*}
\nabla_{\theta}J(\theta) \approx \frac{1}{N} \sum_{i = 1}^{N}  \nabla_{\theta} \log \left(\Pi_{\theta}(\tau_i)\right) R(\tau_i) \\
\theta \leftarrow \theta + \alpha \nabla_{\theta} J(\theta),
\end{align*}
where $\alpha$ is the step-size of the gradient ascent, also known as learning rate.
\section{Proximal policy optimization algorithms} \label{app_PPO}
The policy gradient algorithm suffers from bad sample efficiency and poor convergence. One possible approach to tackle the convergence issue is to use second-order derivative matrices. But such an approach has high computational complexity. Proximal policy optimization (PPO) algorithms are policy gradient algorithms with better sample and computational complexity. The goal of PPO is to estimate a ``trust-region" where one can safely take reasonable steps. For this purpose, PPO maintains two policy networks, one being the current policy network (the policy it is learning) $\Pi_{\theta}\left(a_t \vert s_t\right)$ and a baseline policy network (policy learnt from past experience) $\Pi_{\theta'}\left(a_t \vert s_t\right)$. Let the ratio of these two policies be
\be
r_t(\theta) = \frac{\Pi_{\theta}\left(a_t \vert s_t\right)}{\Pi_{\theta'}\left(a_t \vert s_t\right)}.
\ee
Another concept relevant for understanding PPO is advantage (denoted by $A$), which estimates how good an action is compared to the average action for a particular state. Mathematically, advantage is given by
\be
A^{\pi}(s,a) = Q^{\pi}(s,a) -V^{\pi}(s).
\ee
In the original PPO paper, two different algorithms were presented, i.e. clipped surrogate objective and adaptive KL-penalized objective. We used clipped surrogate objective in this work and so discussed the same here. The clipped PPO objective function clips the estimated advantage function whenever the ratio of new and old policy falls beyond some pre-accepted value. The clipped PPO objective function is given by
\begin{align*}
{\cal{L}}_{\theta^{'}}^{CLIP}(\theta) =&  \mathbb{E}_{\tau \sim \pi^{'}} \left[\sum_{t=0}^{T} [min(r_t(\theta) A_t^{\pi},\right. \\
&\left. clip(r_t(\theta), 1-\epsilon, 1+\epsilon ) A^\pi_t ) ] \right]
\end{align*}
When the ratio of new policy and policy falls outside $(1-\epsilon)$ and $(1+\epsilon)$, the advantage function is assigned the value it would have at the boundaries, i.e. $(1-\epsilon)$ or $(1+\epsilon)$. in the original PPO paper, the value of epsilon was set to $0.2$. As one can expect, $\epsilon$ is a crucial hyperparameter, the value of which tremendously affects the performance of the PPO algorithm.

\section{Details about the variational case}
Hybrid classical-quantum algorithms show immense promise to solve hard problems on near-term quantum devices. In our specific implementation, we choose a set of variational gates that are applied to the initial quantum states of all qubits in state zero. We choose a set of gates that produce real quantum states, as that is sufficient to violate the inequalities maximally. In particular,  we apply a rotation around the $y$ axis $U_n(\theta_n)=\exp(i\theta_n\sigma^y_n)$ on all qubits, where $\sigma^y_n$ is the $y$ Pauli matrix acting on qubit $n$ and $\theta_n$ the rotation angle for qubit $n$. After that, we apply CNOT gates between neighboring qubits $n$ and $n+1$ with periodic boundary condition. These two steps are repeated for $d$ layers. The parameters of the rotational gates are chosen by reinforcement learning. In the last step, another set of rotations are applied.
After creating the quantum state, the set of measurement angles are selected by reinforcement learning. 

The neural network determines the angles as follows: the neural network outputs in each step the mean values for the $n$ angles. The actual chosen angles are sampled from a Normal distribution around these mean values. Then, the angles chosen so far for this epoch are given to the neural network as an input. This procedure is repeated until the angles for the $d$ layers, and the measurement angles are selected. The algorithm is trained to optimize the inequality by using it as a loss function.

\end{document}